\def\Li{{\rm Li}} % The Li function
\documentclass[12pt]{article}
\usepackage{latexsym,amsmath,amsthm}
\newtheorem{lemma}{Lemma}
\date{}

\begin{document}
\pagestyle{myheadings}

\markboth{Closed-form sums$\dots$}{Closed-form sums$\dots$}
\title{Closed-form sums for some perturbation series involving hypergeometric functions}
\author{Nasser Saad$^{\dagger}$ and Richard L. Hall$^{\ddagger}$\\
\\
$^\dagger$Department of Mathematics and Computer Science\\
University of Prince Edward Island\\
550 University Avenue, Charlottetown,\\
Prince Edward Island, Canada C1A 4P3\\
\\
$^\ddagger$Department of Mathematics and Statistics, Concordia University,\\
1455 de Maisonneuve Boulevard West, Montr\'eal, \\
Qu\'ebec, Canada H3G 1M8\\
}
\maketitle

\begin{abstract}
Infinite series of the type $\sum\limits_{n=1}^\infty \frac{(\frac{\alpha}{2})_n}{n}\frac{1}{n!}{}_2F_1(-n,b;\gamma;y)$ are investigated. Closed-form sums are obtained for $\alpha$ a positive integer $\alpha=1,2,3,\dots$. The limiting case of $b\rightarrow \infty$, after $y$ is replaced with $x^2/b$, leads to  $\sum\limits_{n=1}^\infty \frac{(\frac{\alpha}{2})_n}{n}\frac{1}{n!}{}_1F_1(-n,\gamma,x^2)$. This type of series 
appears in the first-order perturbation correction for the wavefunction of the generalized spiked harmonic oscillator Hamiltonian $H=-\frac{d^2}{dx^2}+B x^2 +\frac{A}{x^2}+\frac{\lambda}{x^\alpha}\quad 0\leq x<\infty,\quad \alpha,\lambda > 0,A\geq 0.$ 
These results have immediate applications to perturbation series for the energy and wave function of the spiked harmonic oscillator Hamiltonian  $H=-\frac{d^2}{dx^2}+B x^2 +\frac{\lambda}{x^\alpha}\quad 0\leq x<\infty,\quad \alpha,\lambda > 0.$
\end{abstract}

\bigskip
\newpage
%----------------------------------------------------------------------------------------
%Introduction
%----------------------------------------------------------------------------------------
\section{Introduction}
Recently, Hall et al~{\cite{hsk01,hsk02,hsk03}} have investigated some infinite series of the type 
\begin{equation}\label{E:EI1}
\sum\limits_{n=1}^\infty \frac{(\frac{\alpha}{2})_n}{n}\frac{1}{n!}{}_1F_1(-n;\gamma;x^2),\quad \gamma>\frac{\alpha}{2},\alpha =2,4,\dots
\end{equation}
where
${}_1F_1$ stands for the (Kummer's) confluent hypergeometric function defined in terms of associated Laguerre polynomials by
\begin{equation}\label{E:EI2}
{}_1F_1(-n;\gamma+1;y)=\sum_{k=0}^{n}\frac{(-n)_k}{(\gamma+1)_k}\frac{y^k}{k!}=\frac{n!}{(\gamma+1)_n}L_n^{(\gamma)}(y)
\end{equation}
and $(\gamma)_n,$ the shifted factorial (or {\itshape Pochhammer symbol}), defined by
\begin{equation}\label{EI3}
(\gamma)_0=1,\quad (\gamma)_n=\gamma(\gamma+1)(\gamma+2)\dots (\gamma+n-1)=\frac{\Gamma(\gamma+n)}{\Gamma(\gamma)},\quad n=1,2,\dots.
\end{equation}
where
\begin{equation}\label{E:EI4} 
(-n)_k=\begin{cases}
\frac{(-1)^k n!}{(n-k)!},&\text{ if $0\leq k\leq n$}\\
0,&\text{if $k>n$}.
\end{cases}
\end{equation}

Beside its mathematical relevance as generating functions for Laguerre polynomials~\cite{tos,sm}, the infinite series (\ref{E:EI1}) appears in attempts to develop wavefunction perturbation expansions for the generalized spiked harmonic oscillator Hamiltonian~\cite{hsk04,hsk05,hsk06}
\begin{equation}\label{E:EI5}
H=H_0+\frac{\lambda}{x^\alpha}=-\frac{d^2}{dx^2}+ x^2 +\frac{A}{x^2}+\frac{\lambda}{x^\alpha}\quad 0\leq x<\infty,\quad A\geq 0, \alpha,\lambda > 0,
\end{equation}
which, in turn, has immediate applications in the study of the well-known spiked harmonic oscillator~\cite{hsk01,hsk02,ang} 
\begin{equation}\label{E:EI6}
H=-\frac{d^2}{dx^2}+ x^2 +\frac{\lambda}{x^\alpha}\quad 0\leq x<\infty,\quad \alpha,\lambda > 0.
\end{equation}
In particular, it helps to explain the abnormal behavior of the standard, weak coupling, perturbation theory~\cite{ang} for Hamiltonians (\ref{E:EI6}). In the usual elementary sense the potential $1/x^{\alpha}$ is singular whenever $\alpha > 0.$ However, the value $\alpha = \frac{5}{2}$ marks the division of the problem into two quite distinct regimes: we shall adopt the  convention (following Harrell~\cite{har}) that the term ``singular'' means $\alpha \geq \frac{5}{2};$ some writers such as Klauder~\cite{klau} use instead the term ``supersingular'' for this regime.   Hall et al found~\cite{hsk04,hsk05,hsk06} that for the Hamiltonian (\ref{E:EI5}) defined on the one-dimensional space $(0\leq x<\infty)$ with eigenfunctions satisfying Dirichlet boundary conditions, that is to say, with wavefunctions vanishing at the boundaries, the matrix elements of the operator $x^{-\alpha}$, with respect to the exact solutions of the Gol'dman and Krivchenkov Hamiltonian $H_0$, namely
\begin{equation}\label{E:EI7}
\psi_n(x)=(-1)^n\sqrt{\frac{2(\gamma)_n}{n!\Gamma(\gamma)}}x^{\gamma-\frac{1}{2}}e^{-x^2/2}{}_1F_1(-n,\gamma,x^2)
\end{equation}
with exact eigenenergies
\begin{equation}\label{E:EI8}
E_n=2(2n+\gamma),\quad n=0,1,2,\dots,\quad \gamma=1+\frac{1}{2}\sqrt{1+4A}
\end{equation}
are given explicitly by the following expressions
\begin{equation}\label{E:EI9}
x_{mn}^{-\alpha}=
(-1)^{n+m}
\frac{{(\frac{\alpha}{2})_n}}{(\gamma)_n}
\frac{\Gamma(\gamma-\frac{\alpha}{2})}{\Gamma(\gamma)}
\sqrt{\frac{(\gamma)_n(\gamma)_m}{n!m!}}
{}_3F_{2}(-m,\gamma-\frac{\alpha}{2},1-\frac{\alpha}{2};\gamma,1-n-\frac{\alpha}{2};1)
\end{equation}
and they are valid for all values of the parameters  $\gamma$ and $\alpha$ such that $\gamma>\frac{\alpha}{2}$. Furthermore, the matrix elements of the Hamiltonian (\ref{E:EI5}) are given by
\begin{align}\label{E:EI10}
H_{mn}=2(2n+\gamma)\delta_{mn}+&(-1)^{m+n}\lambda
\sqrt{\frac{(\gamma)_n(\gamma)_m}{n! m!}}
\frac{{\Gamma(\gamma-\frac{\alpha}{2})(\frac{\alpha}{2})_n}}{(\gamma)_n\Gamma(\gamma)}\times\notag\\
&{}_3F_2(-m,\gamma-\frac{\alpha}{2},1-\frac{\alpha}{2};\gamma,1-\frac{\alpha}{2}-n;1).
\end{align}
where $\delta_{mn}$ denotes the Kronecker delta. Of particular interest are the elements
\begin{equation}\label{E:EI11}
H_{0n}= (-1)^{n}\lambda
\sqrt{\frac{(\gamma)_n}{n!}}
\frac{\Gamma(\gamma-\frac{\alpha}{2})}{\Gamma(\gamma)}\frac{(\frac{\alpha}{2})_n}{(\gamma)_n},\quad n\neq 0.
\end{equation}
It is known that the first correction to the wavefunction by means of standard perturbation techniques leads to
\begin{equation}
\psi_0^{(1)}(x)=\sum\limits_{n=1}^{\infty}\frac{H_{0n}}{E_0-E_n}\psi_n(x),\notag
\end{equation}
where $\psi_n(x)$ and $H_{0n}$ are given by (\ref{E:EI7}) and (\ref{E:EI11}), respectively. Thus, the first correction to the wavefunction of the Hamiltonian (\ref{E:EI5}) is given by
\begin{equation}\label{E:EI12}
\psi_0^{(1)}(x)=-\frac{1}{2\sqrt{2}}
\frac{\Gamma(\gamma-\frac{\alpha}{2})}{\Gamma(\gamma)\sqrt{\Gamma(\gamma)}}
x^{\gamma-\frac{1}{2}}e^{-\frac{x^2}{2}}\sum\limits_{n=1}^\infty\frac{(\frac{\alpha}{2})_n}{ n}\frac{1}{n!}{}_1F_1(-n,\gamma,x^2).
\end{equation}
Very little indeed is known about the infinite series which appears in (\ref{E:EI12}) for arbitrary values of $\alpha.$ The first attempt to sum this series, for the very particular case $\alpha=2$, is due to Toscano~\cite{tos} who proved, by means of extensive use of calculus of finite difference~\cite{jor}, and in terms of associated Laguerre polynomials~(\ref{E:EI2}), that
\begin{equation}\label{E:EI13}
\sum\limits_{n=1}^\infty \frac{(n-1)!}{(\gamma)_n}L_n^{(\gamma-1)}(y)= \psi(\gamma)-\log y,\quad\gamma>1
\end{equation}
where $\psi(\gamma)$ is the Digamma function defined by $\psi(\gamma)=\frac{d\log(\Gamma(\gamma))}{d\gamma}$. Another attempt to sum this series  was due to Aguilera-Navarro and Guardiola~\cite{ang}, who reported that they met some serious convergence difficulties even in the case $\alpha=2$ when trying to get the first correction to the wave function. Recently, however, Hall et al gave a systemic method for summing this series in the case of even $\alpha=2,4,6,\dots.$ 
They proved for $\frac{\alpha}{2}= 2+m, m =0,1,2,\dots$ that
\begin{align}\label{E:EI14}
\sum\limits_{n=1}^\infty \frac{(2+m)_n}{n\ n!}&{}_1F_1(-n;\gamma;x^2)=
\psi(\gamma)-\log x^2-\notag\\
&(m+1)\sum\limits_{k=0}^m \frac{(-m)_k}{(k+1)^2}\bigg(-\frac{1}{x^2}\bigg)^k\bigg[L_k^{\gamma-1-k}(x^2)
-\frac{(\gamma-1)}{x^2}L_k^{\gamma-2-k}(x^2)\bigg]
\end{align}
where $L_k^{\gamma}(y)$ defined by~(\ref{E:EI2}), and furthermore, in the case of $\alpha=2$,
\begin{equation}\label{E:EI15}
\sum\limits_{n=1}^\infty \frac{1}{n}{}_1F_1(-n;\gamma;x^2)=
\psi(\gamma)-\log x^2
\end{equation}
In this article, we continue our search for closed-form sums of the perturbation series appearing in (\ref{E:EI12}) by developing more effective methods for the even case and by introducing new methods for the odd integers $\alpha=1,3,5,\dots$. We achieve this by studying a more general class of infinite series involving the classical hypergeometric functions ${}_2F_1$ of Gauss, namely
\begin{equation}\label{E:EI16}
\sum\limits_{n=1}^\infty \frac{(\frac{\alpha}{2})_n}{n\ n!}{}_2F_1(-n,b;\gamma;y),\quad \gamma>\frac{\alpha}{2},\alpha=1,2,\dots
\end{equation}
This leads, as a result of the limiting case of,
\begin{equation}\label{E:EI17}
\lim_{b\rightarrow \infty}{}_2F_1(-n,b;\gamma;\frac{x^2}{b})={}_1F_1(-n;\gamma;x^2)
\end{equation}
to our series in (\ref{E:EI12}). The family of infinite series (\ref{E:EI16}) and (\ref{E:EI1}) are related to the theory of generating functions as well to the theory of the inverse Laplace transform: as a side goal we shall explore these links too. The paper is organized as follows: in Sec. II we demonstrate that the infinite series (\ref{E:EI16}) converges for all $y>0$ and $\gamma>\frac{\alpha}{2}$. Furthermore, we prove that the series (\ref{E:EI16}) has the following integral representation valid for $\gamma>b$
\begin{align}
\sum\limits_{n=1}^\infty \frac{(\frac{\alpha}{2})_n}{n\ n!}{}_2F_1(-n,b;\gamma;y)&=\frac{\alpha}{2}\frac{\Gamma(\gamma)}{\Gamma(b)\Gamma(\gamma-b)}\times\notag\\
&\int\limits_{0}^{1}t^{b-1}(1-t)^{\gamma-b-1}(1-yt)
{}_3F_2(1+\frac{\alpha}{2},1,1;2,2;1-yt)~dt\notag
\end{align}
In order to evaluate the integral just mentioned,  we develop, in Sec. III, analytic expressions of the series ${}_3F_2(1+\frac{\alpha}{2},1,1;2,2;z)$ for arbitrary values of $\alpha$ and $|z|<1$ and we shall prove that
\begin{equation}
z\ {}_3F_2(a+1,1,1; 2, 2; z)=
\frac{1}{a}\int_1^{1-z}\frac{t^a-1}{t^a(1-t)}~dt,\quad a \neq 0.\notag
\end{equation}
Sec. IV is devoted to the even case $\alpha=2,4,6,\dots,m+2, m=0,1,2,\dots$ where we shall prove, for $y>0$, the general formula
\begin{align}
\sum_{n=1}^\infty &\frac{(m+2)_n}{n~n!}{}_2F_1(-n,b;\gamma;y)=\psi(\gamma)-\psi(b)-\log(y)\notag\\
&+(m+1)\sum_{k=0}^m\frac{(-m)_k(1)_k}{(2)_k(2)_k}\bigg[\frac{(\gamma-1)}{(b-1)y}{}_2F_1(-k,2-\gamma;2-b;\frac{1}{y})-{}_2F_1(-k,1-\gamma;1-b;\frac{1}{y})\bigg],\notag
\end{align}
which leads, as we shall demonstrate, to (\ref{E:EI14}) as a special limiting case.
In Sec. V, we shall find closed-form sums for the series (\ref{E:EI16}) for the cases $\alpha=1,3,5,\dots$, and hence we shall prove that
\begin{align}
\sum_{n=1}^\infty \frac{(\frac{1}{2})_n}{n~n!}{}_2F_1(-n,b;\gamma;y)&=2\log(2)+\frac{b}{\gamma}y~{}_3F_2(1,1,1+b;2,1+\gamma;y)\notag\\
&-\frac{2\Gamma(\gamma)\Gamma(b+\frac{1}{2})}{\Gamma(b)\Gamma(\gamma+\frac{1}{2})}\sqrt{y}~{}_3F_2(1,\frac{1}{2},\frac{1}{2}+b;\frac{3}{2},\frac{1}{2}+\gamma;y)\notag
\end{align}
which generate closed form sums for arbitrary odd-integer values $\alpha=1,3,5,\dots$. Closed-form sums for the series (\ref{E:EI1}) then follow as a special limiting cases of the result just mentioned. Finally, in Sec. VI we 
study some applications to the wave function perturbation expansion of the generalized spiked harmonic Hamiltonian: we shall give a complete first-order correction  to the wave function in the case of the radial harmonic oscillator as well as for the spiked harmonic oscillator Hamiltonians. Some other applications to the theory of inverse Laplace transform and generating functions will also be discussed.

It is clear that the functions ${}_1F_1$, ${}_3F_2$, and ${}_2F_1$ , mentioned above, are special cases of the generalized hypergeometric function 
\begin{equation}\label{E:EI18}
{}_pF_{q}(\alpha_1,\alpha_2,\dots,\alpha_p;\beta_1,\beta_2,\dots,\beta_q;z)=
\sum\limits_{k=0}^\infty 
\frac{\prod\limits_{i=1}^p(\alpha_i)_k}{\prod\limits_{j=1}^q(\beta_j)_k}\frac{z^k}{ k!},
\end{equation}
This series converges for all $z$ if $p<q+1$, and for $|z|<1$ if $p=q+1$. It has been assumed that all parameters have real values except for the $\beta_j,j=1,2,\dots,q,$ none of which is equal to zero or to a negative integer. If $p=q+1$, the series is absolutely convergent on the unit circle $|z|=1$ if   
\begin{equation}\label{E:EI19}
{\sum\limits_{j=1}^q \beta_j-\sum\limits_{i=1}^p \alpha_i}>0.
\end{equation}
If one or more $\alpha_i$ are negative integers, the series terminates as a result of (\ref{E:EI4}) and convergence does not enter the discussion. If $p=q$, the series (\ref{E:EI18}) converges for finite $z$.
%---------------------------------------------------------------------------
% 2. Integral Representation
%---------------------------------------------------------------------------
\section{Integral Representation and the Convergence Problem}\medskip

In this section we prove the convergence of the infinite series (\ref{E:EI16}) and develop an  integral representation suitable for our calculations. It is important, however, to notice  for $y=0$, the infinite series (\ref{E:EI16}) for $\alpha\geq 2$ indeed diverges. Since ${}_2F_1(-n,b;\gamma;0)=1$ and $(\frac{\alpha}{2})_{n+1}=\frac{\alpha}{2}(\frac{\alpha}{2}+1)_n$, we have ~\cite{lu01}
\begin{equation}\label{E:EI20}
\sum\limits_{n=1}^\infty \frac{(\frac{\alpha}{2})_n}{n~n!}=\frac{\alpha}{2} \sum\limits_{n=0}^\infty \frac{(1+\frac{\alpha}{2})_n(1)_n(1)_n}{(2)_n(2)_n\ n!}=\frac{\alpha}{2}{}_3F_2(1+\frac{\alpha}{2},1,1;2,2;1)=\psi(1)-\psi(1-\frac{\alpha}{2})
\end{equation}
which is absolutely convergent for $\alpha<2$. Furthermore, for $y = 1$, it is known that
\begin{equation}
{}_2F_1(-n,b;\gamma;1)= \frac{(\gamma-b)_n}{(\gamma)_n}\notag
\end{equation}
as a result of Chu-Vandermonde theorem~\cite{grr}. Thus
\begin{align}\label{E:EI21}
\sum\limits_{n=1}^\infty \frac{(\frac{\alpha}{2})_n}{n}\frac{(\gamma-b)_n}{(\gamma)_n}\frac{1}{n!}&=\frac{\alpha(\gamma-b)}{\gamma} \sum\limits_{n=0}^\infty \frac{(1+\frac{\alpha}{2})_n(\gamma-b+1)_n(1)_n(1)_n}{(\gamma+1)_n(2)_n(2)_n}\frac{1}{n!}\notag\\
&=\frac{\alpha(\gamma-b)}{\gamma}{}_4F_3(1+\frac{\alpha}{2},\gamma-b+1,1,1;\gamma+1,2,2;1)
\end{align}
which is absolutely convergent  for $\frac{\alpha}{2}<1+b$ due to (\ref{E:EI19}). For arbitrary $y>0$, and in order to sum the infinite series (\ref{E:EI16}) for $2\gamma > \alpha>0$, we require a suitable integral representation of the hypergeometric function ${}_2F_1(-n,b;\gamma; y)$ over an appropriate contour, in order to interchange summation with integration and thereby readily conclude the absolute convergence of the series just mentioned. We find the Euler's integral representation~\cite{grr1} 
\begin{equation}
{}_2F_1(a,b;\gamma; y)=\frac{\Gamma(\gamma)}{\Gamma(b)\Gamma(\gamma-b)}
\int\limits_{0}^{1}t^{b-1}(1-t)^{\gamma-b-1}(1-yt)^{-a}~dt\notag
\end{equation}
under the conditions $Re(\gamma)>Re(b)>0$, and $|y|<1$ to be most advantageous for achieving this end. For $a=-n$, these conditions can be relaxed as a result of (\ref{E:EI4}), we have
\begin{equation}\label{E:EI22}
{}_2F_1(-n,b;\gamma; y)=\frac{\Gamma(\gamma)}{\Gamma(b)\Gamma(\gamma-b)}
\int\limits_{0}^{1}t^{b-1}(1-t)^{\gamma-b-1}(1-yt)^{n}dt
\end{equation}
for arbitrary value of $y>0$, which, when substituted into the summation of (\ref{E:EI16}), yields
\begin{equation}\label{E:EI23}
\sum\limits_{n=1}^\infty \frac{(\frac{\alpha}{2})_n}{n\ n!}{}_2F_1(-n,b;\gamma;y)=\frac{\Gamma(\gamma)}{\Gamma(b)\Gamma(\gamma-b)}
\sum\limits_{n=1}^\infty \frac{(\frac{\alpha}{2})_n}{n~n!}
\int\limits_{0}^{1}t^{b-1}(1-t)^{\gamma-b-1}(1-yt)^{n}~dt
\end{equation}
The evaluation of the this last infinite sum, involving integrations over the open interval $(0,1)$, is achieved by examining the summation of the integrand, namely
\begin{equation}
\sum\limits_{n=1}^\infty \frac{(\frac{\alpha}{2})_n}{n~n!}
t^{b-1}(1-t)^{\gamma-b-1}(1-yt)^{n}=t^{b-1}(1-t)^{\gamma-b-1}\sum\limits_{n=1}^\infty \frac{(\frac{\alpha}{2})_n}{n~n!}
(1-yt)^{n}\notag
\end{equation}
and demonstrating that it has an $L_1(0,1)$-majorant. Hence, the existence of such a majorant shall permit us to interchange summation with integration, as result of the Lebesgue Dominated Convergence Theorem. To arrive at such a majorant, we continue by noting that
\begin{align}\label{E:EI24}
\sum\limits_{n=1}^\infty \frac{(\frac{\alpha}{2})_n}{n~n!}
(1-yt)^{n}
&=\frac{\alpha}{2}(1-yt)\sum\limits_{n=0}^\infty \frac{(1+\frac{\alpha}{2})_{n}(1)_n(1)_n} {(2)_n (2)_n}\frac{(1-zt)^{n}}{n!}\notag\\
&=\frac{\alpha}{2}(1-yt){}_3F_2(1+\frac{\alpha}{2},1,1;2,2;1-yt)
\end{align}
which is convergent for $|1-yt|<1$ as a result of (\ref{E:EI19}). Thus
\begin{align}\label{E:EI25}
|\sum\limits_{n=1}^\infty \frac{(\frac{\alpha}{2})_n}{n~n!}
t^{b-1}(1-t)^{\gamma-b-1}(1-yt)^{n}|&=\frac{\alpha}{2}t^{b-1}(1-t)^{\gamma-b-1}|1-yt||{}_3F_2(1+\frac{\alpha}{2},1,1;2,2;1-yt)|\notag\\
&<A(\alpha)t^{b-1}(1-t)^{\gamma-b-1},\quad\quad t\in (0,1)
\end{align}
where the absolute convergence of ${}_3F_2$ and also $|1-yt|<1$ were made use of. The most important aspect of inequality (\ref{E:EI25}) is the appearance of the $L_1(0,1)$-function $t^{b-1}(1-t)^{\gamma-b-1}$, for $\gamma>b$, of variable $t$ majorizing the series
\begin{equation}
\sum\limits_{n=1}^\infty \frac{(\frac{\alpha}{2})_n}{n~n!}
|t^{b-1}(1-t)^{\gamma-b-1}(1-yt)^{n}|.\notag
\end{equation}
This aspect justifies the evaluation of summation (\ref{E:EI23}) by means of the Lebesgue Dominated Convergence Theorem. Thus, we have specifically, for $\gamma>b$, 
\begin{align}\label{E:EI26}
\sum\limits_{n=1}^\infty \frac{(\frac{\alpha}{2})_n}{n\ n!}{}_2F_1(-n,b;\gamma;y)&=\frac{\alpha}{2}\frac{\Gamma(\gamma)}{\Gamma(b)\Gamma(\gamma-b)}\times\notag\\
&\int\limits_{0}^{1}t^{b-1}(1-t)^{\gamma-b-1}(1-yt)
{}_3F_2(1+\frac{\alpha}{2},1,1;2,2;1-yt)~dt
\end{align}
For the computation of this expression, we need analytic expressions for the hypergeometric functions ${}_3F_2(1+\frac{\alpha}{2},1,1;2,2;1-yt):$ these are obtained in the next section.
\medskip
%---------------------------------------------------------------------------
% 3. Analytic expressions for ${}_3F_2(1+\frac{\alpha}{2},1,1;2,2;z)$
%---------------------------------------------------------------------------
\section{Analytic expressions for ${}_3F_2(1+\frac{\alpha}{2},1,1;2,2;z)$}
\medskip
In this section we present some introductory results concerning the analytic sum of  ${}_3F_2(\frac{\alpha}{2}+1,1,1;2,2;z)$ for later use. We start first with an interesting relation between a Gauss function and an elementary function, not usually shown in specialized tables.

%---------
% Lemma 1.
%---------
\medskip
\begin{lemma} 
For $|z|<1$
\begin{equation}\label{E:EI27}
z\ {}_2F_1(a+1,1; 2; z)
= \begin{cases}
-\log(1-z),& \text{if $a=0$}\\
-\frac{1}{a}+\frac{1}{a(1-z)^a},&\text{if $a\neq 0$}.
\end{cases}
\end{equation}
\end{lemma}
\begin{proof} From the series representation of the the hypergeometric function ${}_2F_1$, (\ref{E:EI18}), we have
\begin{align}
{}_2F_1(a+1,1; 2; z) &=\sum_{n=0}^\infty \frac{(a+1)_n(1)_n}{(2)_n}\frac{z^n}{n!}\notag\\
&= \sum_{n=0}^\infty \frac{{(a+1)_n}}{n+1}\frac{z^n}{n!}\notag\\
&=\sum_{n=0}^\infty {(a+1)_n}\frac{z^n}{n!}-\sum_{n=1}^\infty \frac{{n(a+1)_n}}{n+1}
\frac{z^n}{n!}\notag\\
&={}_1F_0(a+1;-;z)-\frac{(a+1)z}{2}  \sum_{n=0}^\infty \frac{{(a+2)_n(2)_n}}{(3)_n}
\frac{z^n}{n!}\notag\\
&={}_1F_0(a+1;-;z)-\frac{(a+1)z}{2}{}_2F_1(a+2,2;3;z)\notag 
\end{align}
However, from the identity
\begin{equation}
\frac{d}{dz}{}_2F_1(a,b;c;z)=\frac{ab}{c}{}_2F_1(a+1,b+1;c+1;z)\notag
\end{equation}
we have
\begin{equation}
{}_2F_1(a+1,1; 2; z) = {}_1F_0(a+1;-;z)-z\frac{d}{dz}{}_2F_1(a+1,1;2;z)\notag
\end{equation}
which yields
\begin{equation}
\frac{d}{dz}\bigg[z\ {}_2F_1(a+1,1;2;z)\bigg]=(1-z)^{-(a+1)}\notag
\end{equation}
where we have used the identity ${}_1F_0(a+1;-;z)=(1-z)^{-(a+1)},\quad |z|<1$. Therefore,
\begin{equation}
z\ {}_2F_1(a+1,1;2;z)=\int_0^z (1-t)^{-(a+1)}dt\notag
\end{equation}
For $a=0$, we have 
$
z\ {}_2F_1(1,1;2;z)=-\log (1-z),\notag
$
while for $a\neq 0$ the result 
\begin{equation}
z\ {}_2F_1(a+1,1;2;z)=-\frac{1}{a} +\frac{1}{a(1-z)^a}\notag
\end{equation}
follows immediately by elementary integration.
\end{proof}
%---------
% Lemma 2.
%---------
\medskip
\begin{lemma} For $|z|<1$
\begin{equation}\label{E:EI28}
z\ {}_3F_2(a+1,1,1; 2, 2; z)=\begin{cases}
\Li_2(z),&\text{if $a=0$}\\
\frac{1}{a}\int_1^{1-z}\frac{t^a-1}{t^a(1-t)}~dt,&\text{if $a\neq 0$}.
\end{cases}
\end{equation}
where $\Li_2(z)$ is the dilogarithmic function defined by
$$\Li_2(z)=-\int_0^z\frac{\log(1-t)}{t}~dt.$$
\end{lemma}
\begin{proof} 
From the series representation (\ref{E:EI18}) of the hypergeometric function ${}_3F_2$, we have
\begin{align}
{}_3F_2(a+1,1,1;2, 2; z) &=\sum_{n=0}^\infty \frac{{(a+1)_n(1)_n(1)_n}}{(2)_n(2)_n}\frac{z^n}
{n!}\notag\\
&= \sum_{n=0}^\infty \frac{{(a+1)_n(1)_n}}{(n+1)(2)_n}\frac{z^n}{n!}\notag\\
&=\sum_{n=0}^\infty \frac{(a+1)_n(1)_n}{(2)_n}\frac{z^n}{n!}-\sum_{n=1}^\infty \frac{{n(a+1)_n(1)_n}}{(n+1)(2)_n}\frac{z^n}{n!}\notag\\
&={}_2F_1(a+1,1;2;z)-\frac{(a+1)z}{4} {}_3F_2(a+2,2,2;3,3;z)\notag
\\
&={}_2F_1(a+1,1;2;z)-z\frac{d}{dz}{}_3F_2(a+1,1,1;2,2;z) \notag
\end{align}
which leads to
\begin{equation}
\frac{d}{dz}\bigg[z\ {}_3F_2(a+1,1,1;2,2;z)\bigg]={}_2F_1(a+1,1;2;z),\notag
\end{equation}
or
\begin{equation}
z\ {}_3F_2(a+1,1,1;2,2;z)=\int_0^z {}_2F_1(a+1,1;2;t)\ dt.\notag
\end{equation}
For $a=0$, and using Lemma 1, we get
\begin{equation}
z\ {}_3F_2(1,1,1;2,2;z)=\int_0^z {}_2F_1(1,1;2;t)\ dt=-\int_0^z\frac{\log(1-t)}{ t}dt=\Li_2(z)\notag
\end{equation}
while for $a\neq 0$, we have by means of Lemma 1 that
\begin{equation}
z\ {}_3F_2(a+1,1,1;2,2;z)=\frac{1}{a}\int_1^{1-z}\frac{t^a-1}{t^a(1-t)}~dt\notag
\end{equation}
as required
\end{proof}
The definite integral in Lemma 2 is easy to evaluate. It follows immediately, for example, in case of $a=1$ that 
\begin{equation}\label{E:EI29}
z\ {}_3F_2(2,1,1;2,2;z)=z\ {}_2F_1(1,1;2;z)=-\int_1^{1-z}\frac{1}{t}dt=-\log(1-z)
\end{equation} 
as expected. For $a=\frac{1}{2}$, 
\begin{equation}\label{E:EI30}
z\ {}_3F_2(\frac{3}{2},1,1;2,2;z)=4\log 2 - 4 \log (1+\sqrt{1-z}),
\end{equation}
for $a=2$,  
\begin{equation}\label{E:EI31}
z\ {}_3F_2(3,1,1;2,2;z)=\frac{1}{2}\frac{z}{1-z}-\frac{1}{2}\log(1-z).
\end{equation}
Similar analytic expressions can be obtain for arbitrary values of $a$. However, for $a = m+2,\quad m=0,1,2,\dots$, the following lemma which can be regarded as a limiting case (as $b\rightarrow 1$) of Luke's reduction formula~\cite{luke} 
\begin{equation}
z\ {}_3F_2(a,b,1;c,2;z)=\frac{(c-1)}{(a-1)(b-1)}\bigg[{}_2F_1(a-1,b-1;c-1;z)-1\bigg],\quad |z|<1,\notag
\end{equation}
can be used. 
%---------
% Lemma 3.
%---------
\medskip
\begin{lemma} For $c\neq 1$, and $|\frac{z}{z-1}|<1$,
\begin{equation}\label{E:EI32}
z\ {}_3F_2(a+1,1,1;c,2;z)=\frac{(c-1)}{a}\bigg[\frac{(c-a-1)}{(c-1)}\bigg(\frac{z}{z-1}\bigg){}_3F_2(c-a,1,1;c,2;\frac{z}{z-1})-\log(1-z)\bigg].
\end{equation}
\end{lemma}
\begin{proof} For the proof of this lemma see Lemma 3 in~\cite{hsk01}
\end{proof}

Although, the right hand side of (\ref{E:EI32}) still in terms of ${}_3F_2$, it is indeed more straightforward to deal with in the case of $a =m+2, m =0,1,2,..$. Indeed for $c=2$, we have 
\begin{equation}
z\ {}_3F_2(m+3,1,1;2,2;z)=\frac{1}{m+2}\bigg[-(m+1)\bigg(\frac{z}{z-1}\bigg){}_3F_2(-m,1,1;2,2;\frac{z}{z-1})-\log(1-z)\bigg]\notag
\end{equation}
where ${}_3F_2(-m,1,1;2,2;\frac{z}{z-1})$ is now a polynomial of degree $m$, therefore
\begin{align}\label{E:EI33}
z\ {}_3F_2(m+3,1,1;2,2;z)=\frac{1}{m+2}&\bigg[-(m+1)\bigg(\frac{z}{z-1}\bigg)
\sum_{n=0}^m\frac{(-m)_n(1)_n(1)_n}{(2)_n(2)_n\ n!}\bigg(\frac{z}{z-1}\bigg)^n\notag\\
&-\log(1-z)\bigg],
\end{align}
which leads, for example, to
\begin{equation}
z\ {}_3F_2(4,1,1;2,2;z)=\frac{1}{3}\bigg[\frac{1}{2}\bigg(\frac{z}{1-z}\bigg)^2-\frac{2z}{1-z}-\log(1-z)\bigg]\notag
\end{equation}
The case of unit argument $z=1$, the hypergeometric function ${}_3F_2(a+1,1,1;2,2;1)$ diverges for $a>1$, while it converges absolutely for $a<1$. With the strict inequality $|z|<1$, the function converges absolutely for arbitrary values of $a$. 

%---------------------------------------------------------------------------
% 4. Closed-Form Sums
%---------------------------------------------------------------------------
\section{Closed-form sums for the even case $\alpha=2,4,6,\dots$}
\medskip
In this section we develope closed form sums for the series (\ref{E:EI16}) for $\alpha=2,4,6,\dots$ which generalize the methods and results of Hall et al~{\cite{hsk01,hsk02,hsk03}}. The odd case $\alpha = 1,3,5,\dots$ will be discussed in the next section.
%---------
% Lemma 4
%---------
\medskip
\begin{lemma} For $\gamma>b$ and $y>0$
\begin{equation}\label{E:EI34}
\sum\limits_{n=1}^\infty\frac{1}{n}{}_2F_1(-n,b;\gamma;y)=
\psi(\gamma)-\psi(b)-\log y
\end{equation}
\end{lemma}
\begin{proof}
From $\alpha=2$, using (\ref{E:EI29}), the integral representation (\ref{E:EI26}) leads to
\begin{align}
\sum\limits_{n=1}^\infty \frac{1}{n}{}_2F_1(-n,b;\gamma;y)&=-\frac{\Gamma(\gamma)}{\Gamma(b)\Gamma(\gamma-b)}
\int\limits_{0}^{1}t^{b-1}(1-t)^{\gamma-b-1}\log(t)
dt\notag\\
&-\frac{\Gamma(\gamma)}{\Gamma(b)\Gamma(\gamma-b)}\log(y)
\int\limits_{0}^{1}t^{b-1}(1-t)^{\gamma-b-1}
dt\notag
\end{align}
The second integral on the right hand side follows immediately from (\ref{E:EI22}) by setting $n=0$ and leads to
\begin{equation}\label{E:EI35}
\int\limits_{0}^{1}t^{b-1}(1-t)^{\gamma-b-1}dt =\frac{\Gamma(b)\Gamma(\gamma-b)}{\Gamma(\gamma)}
\end{equation}
while the first integral can be computed by differentiating (\ref{E:EI35}) with respect to $b$, which leads to the difference of $\int\limits_{0}^{1}t^{b-1}(1-t)^{\gamma-b-1}\log(t)~dt$ and $\int\limits_{0}^{1}t^{b-1}(1-t)^{\gamma-b-1}\log(1-t)~dt$, where the latter can be regarded as the derivative of (\ref{E:EI35}) with respect to $\gamma$. This yields
\begin{equation}
\int\limits_{0}^{1}t^{b-1}(1-t)^{\gamma-b-1}\log(t)
dt=\frac{d}{db}\bigg(\frac{\Gamma(b)\Gamma(\gamma-b)}{\Gamma(\gamma)}\bigg)+\frac{d}{d\gamma}\bigg(\frac{\Gamma(b)\Gamma(\gamma-b)}{\Gamma(\gamma)}\bigg)\notag
\end{equation}
which leads, after some simplification, to (\ref{E:EI34}) as required
\end{proof}

It is clear from (\ref{E:EI34}) that the condition $\gamma>b$ can be removed by analytic continuation, consequentely the special case of the infinite series (\ref{E:EI15}) follows by replacing $y$ with $\frac{x^2}{b}$ in (\ref{E:EI34}) and then taking the limit as $b$ approaches $\infty$, this yields
\begin{equation}
\sum\limits_{n=1}^\infty\frac{1}{n}{}_1F_1(-n;\gamma;x^2)=
\lim\limits_{b\rightarrow \infty}\bigg[\psi(\gamma)-\psi(b)+\log~b-\log x^2\bigg]\notag
\end{equation}
as a result of (\ref{E:EI17}). However, from the integral represention of the Digamma function 
\begin{equation}
\psi(b)=\log(b)+\int\limits_0^\infty [t^{-1}-(1-e^{-t})^{-1}]e^{-bt}~dt,\quad \Re(b)>0\notag
\end{equation}
it follows that $\lim\limits_{b\rightarrow \infty}(\psi(b)-\log~b)=0$ and (\ref{E:EI15}) follows immediately. Similar results for lemma (4) can be carried out for the values of $\alpha=4,6,\dots$ using the integral representation (\ref{E:EI26}) and the analytic expressions for ${}_3F_2$ as developed in Sec. 3. However, we shall adopt a different approach which allows us to obtain a general formula for the sum of the series~(\ref{E:EI16}). First, we generalize Buchholz's identity for Laguerre polynomials~\cite{boch}, 
\begin{equation}\label{E:EI36}
\sum_{n=0}^\infty \frac{(-\nu)_n\Gamma(\gamma+\nu+1)}{ \Gamma(n+\gamma+1)}L_n^{(\gamma)}(y)=\sum_{n=0}^\infty \frac{(-\nu)_n\Gamma(\gamma+\nu+1)}{n! \Gamma(\gamma+1)}{}_1F_1(-n;\gamma+1;y) =y^\nu,
\end{equation} 
for $\gamma+\nu>-1,\ \nu\neq 0,1,2,\dots$.
%---------
% Lemma 5
%---------
\begin{lemma} For $\gamma+\nu>0$, $b+\nu>0$, and $y>0$
\begin{equation}\label{E:EI37}
\sum_{n=0}^\infty \frac{(-\nu)_n}{n!}{}_2F_1(-n,b;\gamma;y)=\frac{\Gamma(\nu+b)}{\Gamma(b)}\frac{\Gamma(\gamma)}{\Gamma(\nu+\gamma)}y^\nu,\
\end{equation}
\end{lemma}
\begin{proof}
From the Laplace transform representation~\cite{slat} of $t^{b-1}{}_1F{}_1(-n;\gamma;yt)$ with unit argument, we have
\begin{equation}
{}_2F_1(-n,b;\gamma;y)=\frac{1}{\Gamma(b)}\int\limits_0^\infty e^{-t}t^{b-1}{}_1F_1(-n;\gamma;yt)~dt,
\notag
\end{equation}
Thus
\begin{align}\label{E:EI38}
\sum_{n=0}^\infty \frac{(-\nu)_n}{n!}{}_2F_1(-n,b;\gamma;y)&=\frac{1}{\Gamma(b)}\int\limits_0^\infty e^{-t}t^{b-1}\bigg[\sum_{n=0}^\infty \frac{(-\nu)_n}{n!}{}_1F_1(-n;\gamma;yt)\bigg]~dt\notag\\
&=\frac{\Gamma(\gamma)}{\Gamma(b)\Gamma(\gamma+\nu)}y^\nu\int\limits_0^\infty e^{-t}t^{\nu+b-1}~dt,\quad \gamma+\nu>0\\
&=\frac{\Gamma(\gamma)\Gamma(b+\nu)}{\Gamma(b)\Gamma(\gamma+\nu)}y^\nu,\quad b+\nu>0\notag
\end{align}
where (\ref{E:EI38}) follows from (\ref{E:EI36}) and the last equality follows from the definition of the Gamma function. 
\end{proof}
With the availability of Lemma 4 and Lemma 5, it is relatively easy, using (\ref{E:EI26}),  to find the closed form sums for the series (\ref{E:EI16}) and consequentely the infinite series (\ref{E:EI1}) for $\alpha = 4,6,\dots$. Indeed, for $\alpha = 4$, we note that $(\frac{\alpha}{2})_n=(2)_n=(1+n)(1)_n$, thus
\begin{equation}
\sum\limits_{n=1}^\infty\frac{(2)_n}{n~n!}{}_2F_1(-n,b;\gamma;y)=
\sum\limits_{n=1}^\infty\frac{1}{n}{}_2F_1(-n,b;\gamma;y)+\sum\limits_{n=1}^\infty {}_2F_1(-n,b;\gamma;y)\notag
\end{equation}
The first series on the right hand side is summable by means of Lemma 4, while for the second series, it is enough to take $\nu=-1$ in the Lemma 5. This leads to
\begin{equation}
\sum\limits_{n=1}^\infty\frac{(2)_n}{n~n!}{}_2F_1(-n,b;\gamma;y)=
\psi(\gamma)-\psi(b)-\log y
+\frac{\gamma-1}{(b-1)y}-1,\quad b>1,\gamma>1\notag
\end{equation}
and consequently
\begin{equation}
\sum\limits_{n=1}^\infty\frac{(2)_n}{n~n!}{}_1F_1(-n;\gamma;x^2)=
\psi(\gamma)-\log x^2
+\frac{\gamma-1}{x^2}-1\notag
\end{equation}
follows by means of (\ref{E:EI17}) and $\lim_{b\rightarrow \infty}(\psi(b)-\log(b))=0$ as mentioned earlier. Similarly, for $\alpha=6$ and $(3)_n=\frac{1}{2}(2+3n+n^2)(1)_n$, we have 
\begin{align}
\sum\limits_{n=1}^\infty\frac{(3)_n}{n~n!}{}_2F_1(-n,b;\gamma;y)=
\sum\limits_{n=1}^\infty\frac{1}{n}{}_2F_1(-n,b;\gamma;y)&+\frac{3}{2}\sum\limits_{n=1}^\infty {}_2F_1(-n,b;\gamma;y)\notag\\
&+\frac{1}{2}\sum\limits_{n=1}^\infty n~{}_2F_1(-n,b;\gamma;y)\notag
\end{align}
Closed form sums for the first and second infinite series on the right-hand side can be found as mentioned above, for the third infinite series it is enough to take $\nu = -2$ in Lemma 5. These results lead to
\begin{equation}\label{E:EI39}
\sum\limits_{n=1}^\infty\frac{(3)_n}{n~n!}{}_2F_1(-n,b;\gamma;y)=\psi(\gamma)-\psi(b)-\log y
+\frac{1}{2}\frac{(\gamma-1)(\gamma-2)}{(b-1)(b-2)}\frac{1}{y^2}+\frac{\gamma-1}{(b-1)y}-\frac{3}{2}
\end{equation}
for $\gamma, b >2$, and consequentely
\begin{equation}\label{E:EI40}
\sum\limits_{n=1}^\infty\frac{(3)_n}{n~n!}{}_2F_1(-n;\gamma;x^2)=\psi(\gamma)-\log x^2
+\frac{1}{2}\frac{(\gamma-1)(\gamma-2)}{x^4}+\frac{\gamma-1}{x^2}-\frac{3}{2}
\end{equation}
by means of (\ref{E:EI17}). For arbitrary values of $\frac{\alpha}{2}=m+2,m=0,1,2,\dots$, a general expressions for the closed form sum of the infinite series (\ref{E:EI16}) can be obtained by means of the following Lemma.
%---------
% Lemma 
%---------
\begin{lemma} For $\gamma>m+2$ and $b>m+2$, where ~$m=0,1,2,\dots$
\begin{align}\label{E:EI41}
\sum_{n=1}^\infty &\frac{(m+2)_n}{n~n!}{}_2F_1(-n,b;\gamma;y)=\psi(\gamma)-\psi(b)-\log(y)\notag\\
&+(m+1)\sum_{k=0}^m\frac{(-m)_k(1)_k}{(2)_k(2)_k}\bigg[\frac{(\gamma-1)}{(b-1)y}{}_2F_1(-k,2-\gamma;2-b;\frac{1}{y})-{}_2F_1(-k,1-\gamma;1-b;\frac{1}{y})\bigg],\
\end{align}
\end{lemma}

\begin{proof} From the integral representation (\ref{E:EI26}) and the analytic expression (\ref{E:EI33}), we have 
\begin{align}
\sum_{n=1}^\infty&\frac{(m+2)_n}{n~n!}{}_2F_1(-n,b;\gamma;y)=
-\frac{\Gamma(\gamma)}{\Gamma(b)\Gamma(\gamma-b)}\int_0^1t^{b-1}(1-t)^{\gamma-b-1}\log(yt)~dt\notag\\
&\frac{(m+1)\Gamma(\gamma)}{\Gamma(b)\Gamma(\gamma-b)}
\sum_{k=0}^m\frac{(-m)_k(1)_k(1)_k}{(2)_k(2)_k~k!}\int_0^1t^{b-1}(1-t)^{\gamma-b-1}\bigg(\frac{1}{yt}-1\bigg)\bigg(1-\frac{1}{yt}\bigg)^k~dt\notag
\end{align}
The first integral on the right hand side is easy to evaluate by a similar proof to Lemma 1 and yields
\begin{equation}
-\frac{\Gamma(\gamma)}{\Gamma(b)\Gamma(\gamma-b)}\int_0^1t^{b-1}(1-t)^{\gamma-b-1}\log(yt)~dt
=\psi(\gamma)-\psi(b)-\log(y)\notag
\end{equation}
The second integral, which we denote it by $I_m$, can be written as (since $(1)_n=n!$)
\begin{align}
I_m&=\frac{(m+1)\Gamma(\gamma)}{\Gamma(b)\Gamma(\gamma-b)}
\sum_{k=0}^m\frac{(-m)_k(1)_k}{(2)_k(2)_k}\frac{1}{y}\int_0^1t^{b-2}(1-t)^{\gamma-b-1}\bigg(1-\frac{1}{yt}\bigg)^k~dt\notag\\
&-\frac{(m+1)\Gamma(\gamma)}{\Gamma(b)\Gamma(\gamma-b)}
\sum_{k=0}^m\frac{(-m)_k(1)_k}{(2)_k(2)_k}\int_0^1t^{b-1}(1-t)^{\gamma-b-1}\bigg(1-\frac{1}{yt}\bigg)^k~dt\notag
\end{align}
Since
\begin{equation}
(1-\frac{1}{yt})^k=\sum\limits_{l=0}^k \frac{(-k)_l}{l!}(\frac{1}{yt})^l,\hbox{ finite number of terms},\notag
\end{equation}
we have
\begin{align}
I_m&=\frac{(m+1)\Gamma(\gamma)}{\Gamma(b)\Gamma(\gamma-b)}
\sum_{k=0}^m\frac{(-m)_k(1)_k}{(2)_k(2)_k}\frac{1}{y}
\sum_{l=0}^k\frac{(-k)_l}{l!}\frac{1}{y^l}\int_0^1t^{b-l-2}(1-t)^{\gamma-b-1}~dt\notag\\
&-\frac{(m+1)\Gamma(\gamma)}{\Gamma(b)\Gamma(\gamma-b)}
\sum_{k=0}^m\frac{(-m)_k(1)_k}{(2)_k(2)_k}
\sum_{l=0}^k\frac{(-k)_l}{l!}\frac{1}{y^l}\int_0^1t^{b-l-1}(1-t)^{\gamma-b-1}~dt\notag
\end{align}
where each of the integral on the right hand side can be evaluate by means of Beta function, this leads to
\begin{align}
I_m&=\frac{(m+1)\Gamma(\gamma)}{\Gamma(b)\Gamma(\gamma-b)}
\sum_{k=0}^m\frac{(-m)_k(1)_k}{(2)_k(2)_k}\frac{1}{y}
\sum_{l=0}^k\frac{(-k)_l}{l!}\frac{1}{y^l}
\frac{\Gamma(b-l-1)\Gamma(\gamma-b)}{\Gamma(\gamma-l-1)}\notag\\
&-\frac{(m+1)\Gamma(\gamma)}{\Gamma(b)\Gamma(\gamma-b)}
\sum_{k=0}^m\frac{(-m)_k(1)_k}{(2)_k(2)_k}
\sum_{l=0}^k\frac{(-k)_l}{l!}\frac{1}{y^k}
\frac{\Gamma(b-k)\Gamma(\gamma-b)}{\Gamma(\gamma-k)}.\notag
\end{align}
Using the identities
\begin{equation}
\Gamma(a-l)=\frac{(-1)^l\Gamma(a)}{(1-a)_l},\quad\quad \Gamma(a-l-1)=\frac{(-1)^l\Gamma(a-1)}{(2-a)_l}\notag
\end{equation}
we obtain by means of the series representaion (\ref{E:EI20}) of ${}_2F_1$ that
\begin{align}
I_m=\frac{(m+1)(\gamma-1)}{(b-1)}&
\sum_{k=0}^m\frac{(-m)_k(1)_k}{(2)_k(2)_k}\frac{1}{y}
{}_2F_1(-k,2-\gamma;2-b;\frac{1}{y})\notag\\
&-(m+1)
\sum_{k=0}^m\frac{(-m)_k(1)_k}{(2)_k(2)_k}
{}_2F_1(-k,1-\gamma;1-b;\frac{1}{y})
\notag
\end{align}
which complete the proof of the Lemma.
\end{proof}
The limit cases 
\begin{equation}
\lim_{b\rightarrow \infty}{}_2F_1(-k,2-\gamma;2-b;\frac{b}{x^2})={}_2F_0(-k,2-\gamma;-;-\frac{1}{x^2})\notag
\end{equation}
and 
\begin{equation}
\lim_{b\rightarrow \infty}{}_2F_1(-k,1-\gamma;1-b;\frac{b}{x^2})={}_2F_0(-k,1-\gamma;-;-\frac{1}{x^2})\notag
\end{equation}
which follows as results of 
\begin{equation}
\lim_{b\rightarrow \infty}\frac{\Gamma(2-b)b^n}{\Gamma(2-b+n)}=(-1)^n,\hbox{ and } \lim_{b\rightarrow \infty}\frac{\Gamma(1-b)b^n}{\Gamma(1-b+n)}=(-1)^n\notag
\end{equation}
respectively, lead us to conclude that
\begin{align}\label{E:EI42}
\sum_{n=1}^\infty &\frac{(m+2)_n}{n~n!}{}_1F_1(-n;\gamma;x^2)=\psi(\gamma)-\log(x^2)\notag\\
&+(m+1)\sum_{k=0}^m\frac{(-m)_k(1)_k}{(2)_k(2)_k}\bigg[\frac{(\gamma-1)}{x^2}{}_2F_0(-k,2-\gamma;-;-\frac{1}{x^2})-{}_2F_0(-k,1-\gamma;-;-\frac{1}{x^2})\bigg],\
\end{align}
which agrees with (\ref{E:EI16}) through the identity~\cite{elna}
\begin{equation}\label{E:EI43}
L_n^{(a-n)}(y)= \frac{(-y)^n}{n!}{}_2F_0(-n,-a;-;-\frac{1}{y}),\quad\hbox{for } n=0,1,2,\dots.
\end{equation}
%---------------------------------------------------------------------------
% 5. Closed-Form Sums II
%---------------------------------------------------------------------------
\section{Closed-form sums for the odd case $\alpha=1,3,5,\dots$}
\medskip
We now turn our attention to the case of $\alpha=1,3,5,\dots$, namely $\frac{\alpha}{2}=m+\frac{1}{2}, m =0,1,2,\dots$. In this case, we don't have, indeed, a concise formula similar to (\ref{E:EI41}), and our results will depend on a recursive approach. For example, in the case of $\alpha=3$, we notice that $(\frac{\alpha}{2})_n=(\frac{3}{2})_n=(1+2n)(\frac{1}{2})_n$ and therefore
\begin{align}\label{E:EI44}
\sum\limits_{n=1}^\infty \frac{(\frac{3}{2})_n}{n~ n!}{}_2F_1(-n,b;\gamma;y)&=\sum\limits_{n=1}^\infty \frac{(1+2n)(\frac{1}{2})_n}{n~ n!}{}_2F_1(-n,b;\gamma;y)\notag\\
&=\sum\limits_{n=1}^\infty \frac{(\frac{1}{2})_n}{n\ n!}{}_2F_1(-n,b;\gamma;y)+2\sum\limits_{n=1}^\infty \frac{(\frac{1}{2})_n}{n!}{}_2F_1(-n,b;\gamma;y)
\end{align}
The second series on the right-hand side is summable by means of Lemma 5 which leads, for the case of $\nu= -\frac{1}{2}$, to
\begin{equation}\label{E:EI45}
\sum_{n=1}^\infty \frac{(\frac{1}{2})_n}{n!}{}_2F_1(-n,b;\gamma;y)=\frac{\Gamma(b-\frac{1}{2})}{\Gamma(b)}\frac{\Gamma(\gamma)}{ \Gamma(\gamma-\frac{1}{2})}\frac{1}{\sqrt{y}}-1,\quad \gamma,b>\frac{1}{2}
\end{equation}
Similarly, in the case of $\alpha = 5$, and since $(\frac{5}{2})_n=(1+\frac{2}{3}n)(\frac{3}
{2})_n$, we have
\begin{align}\label{E:EI46}
\sum\limits_{n=1}^\infty \frac{(\frac{5}{2})_n}{n\ n!}{}_2F_1(-n,b;\gamma;y)&=\sum\limits_{n=1}^\infty \frac{(1+\frac{2}{3}n)(\frac{3}{2})_n}{n\ n!}{}_2F_1(-n,b;\gamma;y)\notag\\
&=\sum\limits_{n=1}^\infty \frac{(\frac{3}{2})_n}{n\ n!}{}_2F_1(-n,b;\gamma;y)+\frac{2}{ 3}\sum\limits_{n=1}^\infty \frac{(\frac{3}{2})_n}{\ n!}{}_2F_1(-n,b;\gamma;y)
\end{align}
The second series on the right-hand side again summable by means of Lemma 5 which leads, in the case $\nu= -\frac{3}{2}$, to
\begin{equation}\label{E:EI47}
\sum_{n=1}^\infty \frac{(\frac{3}{2})_n}{n!}{}_2F_1(-n,b;\gamma;y)=\frac{\Gamma(b-\frac{3}{2})}{\Gamma(b)}\frac{\Gamma(\gamma)}{ \Gamma(\gamma-\frac{3}{2})}\frac{1}{\sqrt{y^3}}-1,\quad \gamma>\frac{1}{2}
\end{equation}
while the first series is summable by means of (\ref{E:EI46}), if of course, we have a closed form sum for the infinite series 
\begin{equation}\label{E:EI48}
\sum\limits_{n=1}^\infty \frac{(\frac{1}{2})_n}{n\ n!}{}_2F_1(-n,b;\gamma;x^2)
\end{equation}
Equivalently, for the cases $\alpha = 7, 9,\dots$ which depend on closed-form sums for the infinite series (\ref{E:EI48}). 
\begin{lemma} For $\gamma>b$ 
\begin{align}\label{E:EI49}
\sum_{n=1}^\infty \frac{(\frac{1}{2})_n}{n~n!}{}_2F_1(-n,b;\gamma;y)&=2\log(2)+\frac{b}{\gamma}y~{}_3F_2(1,1,1+b;2,1+\gamma;y)\notag\\
&-\frac{2\Gamma(\gamma)\Gamma(b+\frac{1}{2})}{\Gamma(b)\Gamma(\gamma+\frac{1}{2})}\sqrt{y}~{}_3F_2(1,\frac{1}{2},\frac{1}{2}+b;\frac{3}{2},\frac{1}{2}+\gamma;y)
\end{align}
\end{lemma}

\begin{proof}
From the integral representation (\ref{E:EI26}) and the analytic expression (\ref{E:EI30}) we have using (\ref{E:EI35})
\begin{equation}
\sum_{n=1}^\infty \frac{(\frac{1}{2})_n}{n~n!}{}_2F_1(-n,b;\gamma;y)=
2\log(2)-\frac{2\Gamma(\gamma)}{\Gamma(b)\Gamma(\gamma-b)}\int_0^1t^{b-1}(1-t)^{\gamma-b-1}\log(1+\sqrt{yt})~dt\notag
\end{equation}
The integral on the right hand side can be performed analytically and  leads to
\begin{align}
\int_0^1t^{b-1}&(1-t)^{\gamma-b-1}\log(1+\sqrt{yt})~dt=-\frac{1}{2}\frac{\Gamma(\gamma-b)}{\Gamma(1+\gamma)\Gamma(\frac{1}{2}+\gamma)}\sqrt{y}
\bigg(\sqrt{y}~\Gamma(b+1)\Gamma(\frac{1}{2}+\gamma)\notag\\
&{}_3F_2(1,1,1+b;2,1+\gamma;y)-2\Gamma(\frac{1}{2}+b)\Gamma(1+\gamma){}_3F_2(1,\frac{1}{2},\frac{1}{2}+b;\frac{3}{2},\frac{1}{2}+\gamma;y)
\bigg)\notag
\end{align}
which the reader can verify using symbolic software such as {\it Maple V.} After some careful simplifications, this leads to the result of the lemma.
\end{proof}
The limit cases 
\begin{equation}
\lim_{b\rightarrow \infty}{}_3F_2(1,1,1+b;2,1+\gamma;\frac{x^2}{b})={}_2F_2(1,1;2,1+\gamma;x^2)\notag
\end{equation}
and 
\begin{equation}
\lim_{b\rightarrow \infty}\frac{\Gamma(\frac{1}{2}+b)}{\sqrt{b}\Gamma(b)}{}_3F_2(1,\frac{1}{2},\frac{1}{2}+b;\frac{3}{2},\frac{1}{2}+\gamma;\frac{x^2}{b})={}_2F_2(1,\frac{1}{2};\frac{3}{2},\frac{1}{2}+\gamma;
x^2)\notag
\end{equation}
leads us to conclude for $\alpha=1$
\begin{align}\label{E:EI50}
\sum_{n=1}^\infty \frac{(\frac{1}{2})_n}{n~n!}{}_1F_1(-n;\gamma;x^2)=2\log(2)&+\frac{x^2}{\gamma}~{}_2F_2(1,1;2,1+\gamma;x^2)\notag\\
&-\frac{2x\Gamma(\gamma)}{\Gamma(\gamma+\frac{1}{2})}{}_2F_2(1,\frac{1}{2};\frac{3}{2},\frac{1}{2}+\gamma;x^2),\quad \hbox{for }\gamma>\frac{1}{2}.
\end{align} 
This can be used now to obtain a closed form expressions for the infinite series (\ref{E:EI1}) for $\alpha=3,5,\dots$, indeed we may now conclude using (\ref{E:EI45}) and (\ref{E:EI47}), that
for $\alpha = 3$
\begin{align}\label{E:EI51}
\sum_{n=1}^\infty \frac{(\frac{3}{2})_n}{n~n!}&{}_1F_1(-n;\gamma;x^2)=2\log(2)+\frac{x^2}{\gamma}~{}_2F_2(1,1;2,1+\gamma;x^2)\notag\\
&-\frac{2x\Gamma(\gamma)}{\Gamma(\gamma+\frac{1}{2})}{}_2F_2(1,\frac{1}{2};\frac{3}{2},\frac{1}{2}+\gamma;x^2)+\frac{2\Gamma(\gamma)}{ \Gamma(\gamma-\frac{1}{2})}\frac{1}{x}-2,\quad \hbox{for }\gamma>\frac{3}{2}.
\end{align}
Similarly, for $\alpha=5$
\begin{align}\label{E:EI52}
\sum_{n=1}^\infty &\frac{(\frac{5}{2})_n}{n~n!}{}_1F_1(-n;\gamma;x^2)=2\log(2)+\frac{x^2}{\gamma}~{}_2F_2(1,1;2,1+\gamma;x^2)\notag\\
&-\frac{2x\Gamma(\gamma)}{\Gamma(\gamma+\frac{1}{2})}{}_2F_2(1,\frac{1}{2};\frac{3}{2},\frac{1}{2}+\gamma;x^2)+\frac{2\Gamma(\gamma)}{ \Gamma(\gamma-\frac{1}{2})}\frac{1}{x}+\frac{2\Gamma(\gamma)}{3\Gamma(\gamma-\frac{3}{2})}\frac{1}{x^3}-\frac{8}{3},\quad \hbox{for }\gamma>\frac{5}{2}.
\end{align}
Corresponding expressions follows in a similar fashion for $\alpha=7,9,\dots$.
%---------------------------------------------------------------------------
% 6. Applications
%---------------------------------------------------------------------------
\section{Applications}
\medskip
In this section, we shall discuss some applications for the results developed in Sec. 4 and Sec. 5. For the  Hamiltonian (\ref{E:EI5}) and (\ref{E:EI6}), it is known for the case of $\alpha<\frac{5}{2}$, the potentials are non-singular, and one should expect that the standard perturbation theory could be applied without any difficulties. This is the case for the eigenvalue expansions, where the resulting expansion for the energy is easily obtained and agrees with numerically determined eigenvalues for small $\lambda$. Indeed, It has been shown by Hall et al~\cite{hsk03} that the eigenvalue expansions, for the ground state energy, are given by
\begin{equation}\label{E:EI53}
E_0(\alpha,\gamma)=2\gamma+\frac{\Gamma(\gamma-\frac{\alpha}{2})}{\Gamma(\gamma)}\lambda-\frac{\alpha^2}{16\gamma}\frac{\Gamma^2(\gamma-\frac{\alpha}{2})}
{\Gamma^2(\gamma)} \lambda^2{}_4F_3(1,1,\frac{\alpha}{2}+1,\frac{\alpha}{2}+1;\gamma+1,2,2;1)+\dots
\end{equation}
and valid for all $\alpha<\gamma+1$, where $\gamma=1+\frac{1}{2}\sqrt{1+4A}$. In particular, for $\alpha=1$ and $\gamma>\frac{1}{2}$
\begin{equation}
E_0(1,\gamma)=2\gamma+\frac{\Gamma(\gamma-\frac{1}{2})}{\Gamma(\gamma)}\lambda-\frac{1}{16\gamma}\frac{\Gamma^2(\gamma-\frac{1}{2})}
{\Gamma^2(\gamma)} \lambda^2{}_4F_3(1,1,\frac{3}{2},\frac{3}{2}+1;\gamma+1,2,2;1)+\dots\notag
\end{equation}
which leads for the spiked hamonic oscillator Hamiltonian, i.e. $\gamma=\frac{3}{2}$ or $A=0$, to
\begin{equation}
E_0(1,\gamma)=3+1.128~379~167\lambda-0.077~890~972\lambda^2+\dots\notag
\end{equation}
The eigenfunction expansions, however, cannot be always obtained with the same ease, nor do they have a similar form. To illustrate this point, the first-order expansions of perturbed wavefunctions, for $x>0$, can be obtained using (\ref{E:EI12}) and the results of Sec. 4 and Sec. 5 by the following expressions: 
for $\alpha = 1$,
\begin{align}\label{E:EI54}
\psi_0^{(1)}(x)=-\frac{1}{2\sqrt{2}}
\frac{\Gamma(\gamma-\frac{1}{2})}{\Gamma(\gamma)\sqrt{\Gamma(\gamma)}}
x^{\gamma-\frac{1}{2}}&e^{-\frac{x^2}{2}}
\bigg[2\log(2)+\frac{x^2}{\gamma}~{}_2F_2(1,1;2,1+\gamma;x^2)\notag\\
&-\frac{2x\Gamma(\gamma)}{\Gamma(\gamma+\frac{1}{2})}{}_2F_2(1,\frac{1}{2};\frac{3}{2},\frac{1}{2}+\gamma;x^2)\bigg],\quad \hbox{for }\gamma>\frac{1}{2}
.
\end{align}
For $\alpha=2$
\begin{equation}\label{E:EI55}
\psi_0^{(1)}(x)=
\frac{1}{\sqrt{2}}
\frac{1}{(\gamma-1)\sqrt{\Gamma(\gamma)}}
x^{\gamma-\frac{1}{2}}e^{-\frac{x^2}{2}}\bigg[\log x-\frac{1}{2}\psi(\gamma)\bigg],\quad \hbox{for } \gamma>1.
\end{equation}
For $\alpha=3$
\begin{align}\label{E:EI56}
\psi_0^{(1)}(x)&=-\frac{1}{2\sqrt{2}}
\frac{\Gamma(\gamma-\frac{3}{2})}{\Gamma(\gamma)\sqrt{\Gamma(\gamma)}}
x^{\gamma-\frac{1}{2}}e^{-\frac{x^2}{2}}
\bigg[2\log(2)+\frac{x^2}{\gamma}~{}_2F_2(1,1;2,1+\gamma;x^2)\notag\\
&-\frac{2x\Gamma(\gamma)}{\Gamma(\gamma+\frac{1}{2})}{}_2F_2(1,\frac{1}{2};\frac{3}{2},\frac{1}{2}+\gamma;x^2)+\frac{2\Gamma(\gamma)}{ \Gamma(\gamma-\frac{1}{2})}\frac{1}{x}-2\bigg],\quad \hbox{for }\gamma>\frac{3}{2}.
\end{align}
For $\alpha=4$
\begin{equation}\label{E:EI57}
\psi_0^{(1)}(x)=
\frac{1}{2\sqrt{2}}
\frac{x^{\gamma-\frac{1}{2}}e^{-\frac{x^2}{2}}}{(\gamma-2)(\gamma-1)\sqrt{\Gamma(\gamma)}}
\bigg[\log(x^2)-\psi(\gamma)-\frac{\gamma-1}{x^2}+1\bigg],\quad \hbox{for } \gamma>2.
\end{equation}
and similarly for $\alpha=5,6,7,\dots$. In these expressions, the coefficient of the exponential term is no longer a polynomial in $x$. These are quite surprising results and demonstrate clearly  that the forms of the perturbation series for eigenvalues and wave functions are very different.

It is clear with $A=l(l+1)$, where $l=0,1,2,\dots$ is a quantum number, the Hamiltonian~(\ref{E:EI5}) can be regarded as the perturbation of the radial harmonic oscillator, while for $A=(l+\frac{1}{2}(N-1))(l+\frac{1}{2}(N-3))$ it can be regarded as perturbation of the harmonic oscillator in $N$-dimensions, in either case (\ref{E:EI53})-(\ref{E:EI56}) provide the first-order perturbation correction to the wavefunctions for the cases of $\alpha=1,2,3,\dots$.
There is another important point that follows from the properties of Pochhammer symbol $(a+1)_n=\frac{a+n}{a}(a)_n$ and Lemma 5 in Sec. 4: closed-form sums for the infinite series (\ref{E:EI16}), and consequentely (\ref{E:EI1}) for arbitrary $\alpha>0$ and $\alpha\neq 1,2,3,\dots$ depend only on closed-form sums for series in the case of $0<\alpha<1$. Although we know in this case the potential is non-singular and the Hamiltonian has accurate eigenvalue expansions that agree with the numerical results, the eigenfunction expansions, still however, have non-polynomial expansions.

A side outcome of this work is that our results, especially (\ref{E:EI26}), lead us to obtain some new formulas for the Laplace inverse transform that are not mentioned in the known tables dealing with hypergeometric functions~\cite{rk}. Hall et al~\cite{hsk01} introduced the following Laplace inverse transform representation of the infinite series~(\ref{E:EI1})
\begin{equation}\label{E:EI58}
\sum\limits_{n=1}^\infty\frac{(\frac{\alpha}{2})_n}{n~n!}{}_1F_1(-n,\gamma,x^2)=
\frac{\alpha}{2}\frac{\Gamma(\gamma)}{2\pi i}
\int\limits_{c-i\infty}^{c+i\infty}e^{s}s^{-\gamma}
(1-\frac{x^2}{s})\ {}_3F_2(1,1,1+\frac{\alpha}{2};2,2;1-\frac{x^2}{s})\ ds.
\end{equation} 
The integral representation (\ref{E:EI26}) allow us to conclude that the Laplace inverse transform of $s^{-\gamma}
(1-\frac{x^2}{s})\ {}_3F_2(1,1,1+\frac{\alpha}{2};2,2;1-\frac{x^2}{s})$ with unit argument is
\begin{align}\label{E:EI59}
{\cal L}^{-1}\bigg(s^{-\gamma}
&(1-\frac{x^2}{s})\ {}_3F_2(1,1,1+\frac{\alpha}{2};2,2;1-\frac{x^2}{s})\bigg)=\lim_{b\rightarrow \infty}\frac{1}{\Gamma(b)\Gamma(\gamma-b)}\times \notag\\
&\int\limits_{0}^{1}t^{b-1}(1-t)^{\gamma-b-1}(1-\frac{x^2t}{b})
{}_3F_2(1+\frac{\alpha}{2},1,1;2,2;1-\frac{x^2t}{b})~dt
\end{align}
which leads, for example, in the case of $\alpha=2$, to the known formula
\begin{equation}\label{E:EI60}
{\cal L}^{-1}\bigg(s^{-\gamma}(\log s-\log x^2)\bigg)=\frac{1}{\Gamma(\gamma)}[\psi(\gamma)-\log x^2]
\end{equation}
There are more interesting formulas for Laplace inverse transforms which are not usually mentioned in Tables: for $\gamma>\frac{1}{2}$ and $x>0$
\begin{align}\label{E:EI61}
{\cal L}^{-1}\bigg(s^{-\gamma}\log(\sqrt{s}+x)\bigg)=\frac{1}{2}\frac{\psi(\gamma)}
{\Gamma(\gamma)}
+\frac{x}{\Gamma(\gamma+\frac{1}{2})}
&{}_2F_2(1,\frac{1}{2};\frac{3}{2},\frac{1}{2}+\gamma;x^2)\notag\\-\frac{x^2}{2\Gamma(\gamma+1)}{}_2F_2(1,1;2,1+\gamma;x^2).
\end{align}
Indeed, from (\ref{E:EI58}), we have in the case of $\alpha=\frac{1}{2}$, using (\ref{E:EI30}), that
\begin{align}\label{E:EI62}
\sum\limits_{n=1}^\infty\frac{(\frac{1}{2})_n}{n~n!}{}_1F_1(-n,\gamma,x^2)&=
\frac{1}{2}\frac{\Gamma(\gamma)}{2\pi i}
\int\limits_{c-i\infty}^{c+i\infty}e^{s}s^{-\gamma}
(1-\frac{x^2}{s})\ {}_3F_2(\frac{3}{2},1,1;2,2;1-\frac{x^2}{s})ds\notag\\
&= \frac{\Gamma(\gamma)}{2\pi i}
\int\limits_{c-i\infty}^{c+i\infty}e^{s}s^{-\gamma}\bigg(2\log 2 -2\log(1+\frac{x}{ \sqrt{s}})\bigg)
ds\notag\\
&=\frac{\Gamma(\gamma)}{2\pi i}
\int\limits_{c-i\infty}^{c+i\infty}e^{s}s^{-\gamma}\bigg(2\log 2 -2\log(\sqrt{s}+x)+\log s\bigg)ds\notag\\
&=2\log 2+\psi(\gamma)-\frac{2\Gamma(\gamma)}{2\pi i}
\int\limits_{c-i\infty}^{c+i\infty}e^{s}s^{-\gamma}\log(\sqrt{s}+x)~ds,
\end{align}
where the last equality follows with aid of the identities
\begin{equation}
[\Gamma(\gamma)]^{-1}= \frac{1}{2\pi i} \int\limits_{c-i\infty}^{c+i\infty}e^{t}s^{-\gamma}ds,
\hbox{ and }
\frac{\psi(\gamma)}{\Gamma(\gamma)}=\frac{1}{ 2\pi i}  \int\limits_{c-i\infty}^{c+i\infty}e^{s}s^{-\gamma}\log(s)ds,\quad\quad \Re(\gamma)>0, c>0,\notag
\end{equation}
By comparing (\ref{E:EI62}) with (\ref{E:EI50}), the formula (\ref{E:EI61}) follows. There is independent confirmation for this result and consequently an alternative proof of the important summation~(\ref{E:EI50}). Let us denote the integral representation in (\ref{E:EI62}) by
\begin{equation}\label{E:EI63}
f_\gamma(x)=\frac{1}{2\pi i}
\int\limits_{c-i\infty}^{c+i\infty}e^{s}s^{-\gamma}\log(\sqrt{s}+x)~ds
\end{equation}
For this particular representation it is clear that $f_\gamma(0)=\frac{\psi(\gamma)}{2\Gamma(\gamma)}$. By differentiating (\ref{E:EI63}) with respect to $x$, we get
\begin{align}\label{E:EI64}
\frac{d}{dx}f_\gamma(x)&=\frac{1}{2\pi i}\int\limits_{c-i\infty}^{c+i\infty}e^{s}s^{-\gamma-\frac{1}{2}}(1-\frac{x^2}{s})^{-1}~ds-\frac{x}{2\pi i}\int\limits_{c-i\infty}^{c+i\infty}e^{s}s^{-\gamma-1}(1-\frac{x^2}{s})^{-1}~ds\notag\\
&=\frac{1}{\Gamma(\gamma+\frac{1}{2})}{}_1F_1(1;\frac{1}{2}+\gamma;x^2)-\frac{x}{\Gamma(\gamma+1)}{}_1F_1(1;1+\gamma;x^2),\quad \gamma>-1
\end{align}
where we have used the identity~\cite{slat}
\begin{equation}
\frac{t^{\nu-1}}{\Gamma(\nu)}{}_1F_1(a;\nu;kt)=\frac{1}{2\pi i}\int\limits_{c-i\infty}^{c+i\infty}e^{st}s^{-\nu}(1-\frac{k}{s})^{-a}~ds,\quad \nu>0,s>k\notag
\end{equation}
By integrating (\ref{E:EI64}), using 
\begin{equation}
\int_0^x{}_1F_1(1;\frac{1}{2}+\gamma;t^2)~dt=x~{}_2F_2(1,\frac{1}{2};\frac{3}{2},\frac{1}{2}+\gamma;x^2)\notag
\end{equation}
and
\begin{equation}
\int_0^x t{}_1F_1(1;1+\gamma;t^2)~dt=\frac{x^2}{2}~{}_2F_2(1,1;2,1+\gamma;x^2)\notag
\end{equation}
we get
\begin{equation}
f_\gamma(x)=\frac{\psi(\gamma)}{2\Gamma(\gamma)}+\frac{x}{2\Gamma(\gamma+\frac{1}{2})}{}_2F_2(1,\frac{1}{2};\frac{3}{2},\frac{1}{2}+\gamma;x^2)-\frac{x^2}{2\Gamma(\gamma+1)}{}_2F_2(1,1;2,1+\gamma;x^2)\notag
\end{equation}
which proves (\ref{E:EI61}) and consequently confirms (\ref{E:EI50}). Finally, we mention two identities, among many, that follow easily from our work. It is known~\cite{sm}
\begin{equation}
\sum_{n=0}^\infty\frac{(-\lambda)_n}{n!}{}_2F_1(-n,\alpha;\beta;z)t^n=(1-t)^{\lambda}{}_2F_1(-\lambda,\alpha;\beta;-\frac{zt}{1-t}),\quad |t|<1,\notag
\end{equation}
we may now add, as a result of Lemma 5, that
\begin{equation}\label{E:EI65}
\sum_{n=0}^\infty\frac{(-\lambda)_n}{n!}{}_2F_1(-n,\alpha;\beta;z)=\frac{\Gamma(\lambda+\alpha)\Gamma(\beta)}{\Gamma(\alpha)\Gamma(\lambda+\beta)}z^\lambda,\
\end{equation}
Furthermore, by setting $\gamma=b$ in (\ref{E:EI49}) and for $0<y<1$, we have the following identity
\begin{equation}\label{E:EI66}
{}_2F_1(\frac{1}{2},1;\frac{3}{2};y)=\frac{1}{\sqrt{y}}\log\bigg(\frac{1+\sqrt{y}}{\sqrt{1-y}}\bigg).
\end{equation}
\newpage
%---------------------------------------------------------------------------
% 7. Conclusion
%---------------------------------------------------------------------------
\section{Conclusion}
\medskip
This paper reports results in an on-going study of the spectrum of an oscillator perturbed by a 
term of the form $\lambda/x^{\alpha}.$ This problem exhibits particularly interesting features when it is ``singular'', that is to say when $\alpha \geq \frac{5}{2}.$  In this regime, standard perturbation theory breaks down in the sense that the perturbation is not turned off smoothly as $\lambda\rightarrow 0.$  We use as an un-perturbed basis problem the soluble case $\alpha = 2.$  Although this is is not singular it nevertheless provides a basis that is `closer'
to the original singular problem than would be the eigenfunctions, say, of an unperturbed oscillator.  The perturbation series generated in this way, both for the eigenvalues and for the eigenfunctions, lead us in turn to summation problems of special series of hypergeometric functions.  We are happy to report that, by the present work, all the difficulties of summing these series have now been overcome for the case where $\alpha$ is an integer.  Consequently we are able to provide explicit formulas for the coefficients in the original perturbation series.       

\bigskip
\section*{Acknowledgment}
\medskip Partial financial support of this work under Grant No. GP3438
from the 
Natural Sciences and Engineering Research Council of Canada is gratefully 
acknowledged by one of us [RLH].

\end{document}